\documentclass[letterpaper,twocolumn,10pt]{article}
\usepackage{usenix-2020-09}

\usepackage{amsmath}
\usepackage{tikz}
\usepackage{pgfplots}
\usetikzlibrary{arrows,shadows}
\usepackage{pgf-umlsd}
\usepackage{adjustbox}
\usepackage{array}
\usepackage{cleveref}
\usepackage{xspace}
\usepackage{wasysym}
\usepackage{xurl}

\usepackage{enumitem}
\usepackage{filecontents}
\usepackage{soul}

\newcommand{\etal}{\textit{et al}.}
\newcommand{\ie}{\textit{i}.\textit{e}.,~}
\newcommand{\eg}{\textit{e}.\textit{g}.,~}
\newcommand{\A}{$\mathcal{A}$\xspace}

\renewenvironment{quote}
              {\list{}{\listparindent=0.2em
                        \itemindent=\listparindent
                        \leftmargin=\parindent
                        \rightmargin=\parindent
                        \topsep=0.2em
                        }%
                \item\relax}
              {\endlist}

\newcommand*{\sysname}[1][]{\textit{InfoGuard}{\if\relax\detokenize{#1}\relax\else\textit{#1}\fi}\xspace}

\newcommand{\sysg}{\textit{GUI}_{IG}}

\usepackage{listings}

\definecolor{lightgray}{rgb}{.9,.9,.9}
\definecolor{darkgray}{rgb}{.4,.4,.4}
\definecolor{purple}{rgb}{0.65, 0.12, 0.82}

\lstdefinelanguage{JavaScript}{
  keywords={typeof, new, true, false, catch, function, return, null, catch, switch, var, if, in, while, do, else, case, break},
  keywordstyle=\color{blue}\bfseries,
  ndkeywords={class, export, boolean, throw, implements, import, this},
  ndkeywordstyle=\color{darkgray}\bfseries,
  identifierstyle=\color{black},
  sensitive=false,
  comment=[l]{//},
  morecomment=[s]{/*}{*/},
  commentstyle=\color{purple}\ttfamily,
  stringstyle=\color{red}\ttfamily,
  morestring=[b]',
  morestring=[b]"
}

\begin{document}

\date{}

\title{\sysname: A Design and Usability Study of User-Controlled Application-Independent Encryption for Privacy-Conscious Users}


\author{{Tarun Kumar Yadav, Austin Cook, Justin Hales, Kent Seamons} \\
Brigham Young University}




\maketitle
\begin{abstract}

Billions of secure messaging users have adopted end-to-end encryption (E2EE). Nevertheless, challenges remain. Most communication applications do not provide E2EE, and application silos prevent interoperability. Our qualitative analysis of privacy-conscious users' discussions of E2EE on Reddit reveals concerns about trusting client applications with plaintext, lack of clear indicators about how encryption works, high cost to switch apps, and concerns that most apps are not open source.
We propose \sysname, a system enabling E2EE for user-to-user communication in any application. \sysname allows users to trigger encryption on any textbox, even if the application does not support E2EE. \sysname encrypts text before it reaches the application, eliminating the client app's access to plaintext. \sysname also incorporates visible encryption to make it easier for users to understand that their data is being encrypted and give them greater confidence in the system's security. The design enables fine-grained encryption, allowing specific sensitive data items to be encrypted while the rest remains visible to the server. Participants in our user study found \sysname usable and trustworthy, expressing a willingness to adopt it. 

\end{abstract}



\section{Introduction}

TLS dominates the encryption landscape for encrypting data during transmission to protect it from eavesdroppers and active attackers. However, attackers can access sensitive information by compromising the endpoints of a connection. Thus, users must trust the client application and server with sensitive data. 

End-to-end encryption (E2EE) prevents server access by having two clients communicate through a server and encrypt and decrypt the sensitive data only at the client endpoints.
The most prominent use of E2EE is the secure messaging apps used by billions of users. Most E2EE applications are siloed, requiring users to adopt the same client application to communicate securely. 
Furthermore, users must trust the client-side app because sensitive data is given to the app before it is encrypted. Trusting the client but not the server is contradictory because the server typically supplies the client app. 
The threat model of trusting the client may concern privacy-conscious users, such as people who adopted secure email due to privacy concerns and a lack of trust in big tech~\cite{289542}. 

To gain insight into the perceptions and expectations of E2EE for privacy-conscious users, we conducted a qualitative analysis of discussions on three prominent subreddits: "privacy," "privacyToolsIO," and "privacyGuides." These subreddits have a combined membership of 1.56 million users. From a pool of 5,298 relevant posts that discuss communication privacy, we randomly selected and coded 210 posts to identify common themes and concerns.



The results from our Reddit analysis show that application siloing is a problem. Applications, not users, decide when to encrypt data. Currently, most popular applications, such as Email, Slack, Microsoft Teams, and Discord, still use TLS, making it difficult for users to secure their communication. Moreover, even when the applications provide E2EE, their client apps have access to plaintext, which could lead to unintended leakage due to bugs or intentional breaches. Furthermore, users often do not trust E2EE applications, already mentioned by~\cite{abu2018exploring}, as they cannot visibly distinguish between plaintext communication, TLS encryption, and E2EE. 
Because encryption is not visible, users must rely on the app's description to understand how it secures data. 


Based on the Reddit analysis, we designed and implemented two variants of \sysname that provided user-controlled, application-independent encryption without trusting the client and server application. Furthermore, \sysname provides fine-grained control over what is encrypted in an application. The two \sysname provide the same security \& privacy properties but different UX, which we test through a usability study.


Both \sysname monitor the user's keypresses directly from the keyboard driver for a trigger to start encryption mode. Users can request encryption on any textbox on any application by pressing a keyboard shortcut (ctrl-alt-e) to activate encryption mode, even if the application does not support E2EE. Once encryption is activated, \sysname receives an exclusive lock on the keyboard output and encrypts it using the Signal protocol before passing ciphertext to the windowing system, such as Wayland, which is responsible for passing the keypresses to applications.
\sysname{v1} opens a new secure graphical user interface ($\sysg$) window to display the plaintext as the user enters it, but the data is not accessible to any other user-level application. \sysname{v2} opens a new $\sysg$ window where the user types and the encrypted text is automatically passed to the target application.
The users have visual evidence that the application is receiving only ciphertext.
When the recipient receives the ciphertext in the application, the user needs to have \sysname installed to read the plaintext. The user can select the ciphertext, and enter the decryption shortcut to decrypt the message. \sysname then generates a $\sysg$ on the recipient's screen that displays the decrypted text.



\sysname is a novel approach to E2EE by providing a centralized application-independent E2EE architecture that eliminates the client app's access to plaintext, incorporates visible encryption to provide users with increased confidence in the system's security, and fine-grained encryption to encrypt only the sensitive parts of a message. 

In summary, our contributions include:  (1) an analysis of privacy-conscious Reddit users discussing their concerns and preferences of communication platforms and E2EE,
(2) the design of \sysname, an architecture for user-controlled, application-independent E2EE,
(3) proof-of-concept implementations of \sysname{v1} and \sysname{v2}, along with a performance analysis,
(4) a comparative analysis of \sysname to other E2EE tools, 
(5) a user study demonstrating the usability and trust in \sysname,
and (6) a discussion on how \sysname can be used by app developers to provide E2EE in their apps.

\section{Related Work and Background}

\subsection{Secure Communication}

PGP~\cite{zimmermann1995official} is an early secure email protocol having asynchronous and high-latency properties. Many PGP-based tools made it available to Internet users~\cite{PGP_tool, GPGTools}. Unger \etal~\cite{unger2015sok} present a systematization of knowledge (SoK) paper on secure communication tools. Later, Clark \etal~\cite{clark2021sok} present an SoK that analyzes the stakeholders of email communication and their different priorities regarding security goals, utility, deployability, and usability. 


OTR~\cite{otr} presents the first secure messaging protocol that works for low latency and has preferable security properties such as perfect forward secrecy and deniability. There are many other secure messaging protocols~\cite{borisov2004off, alexander2007improved, bian2007off, stedman2008user, goldberg2009multi, liu2013improved, signal_app} based on OTR that focus on improving security properties, usability, or features like group messaging. 

\paragraph{Standalone encryption applications} There exists many systems that provide application-specific encryption like PGP~\cite{PGP_tool} (including S/MIME~\cite{ramsdell2010secure}), TextSecure~\cite{textsecure}, FlyByNight~\cite{lucas2008flybynight}, Gibberbot~\cite{Gibberbot}, and SafeSlinger~\cite{farb2013safeslinger}. Previous research~\cite{abu2017obstacles} shows that fragmented user bases are a major obstacle to adopting security protocols. Therefore, we propose a system designed to work with all existing applications without requiring any modifications and to enable user control of encryption. 

\textit{Messaging applications: }
Secure messaging applications use many different protocols to provide E2EE. One family of applications uses the Signal protocol~\cite{signalProtocol, cohn2017formal}, hereafter referred to as Signal. These include the Signal app, WhatsApp~\cite{whatsappencryption}, Facebook Messenger~\cite{messengersecret}, Skype~\cite{skypeprivate}, and Riot\footnote{Riot uses Olm, an implementation of the Signal Double Ratchet algorithm, for one-to-one encrypted communication (https://gitlab.matrix.org/matrix-org/olm/blob/master/docs/olm.md).}, all of which directly use the Signal protocol, as well as Wire~\cite{wiresecurity}, and Viber~\cite{viberencryption}, which use proprietary implementations but follow the same concepts.
Another family of applications (\eg, iMessage, Threema, and Wickr) uses a proprietary protocol that bootstraps encryption by exchanging public keys using a central server, similar to the initialization used by Signal.

\paragraph{Browser-based encryption systems: }Fahl et al.~\cite{fahl2012confidentiality} designed a Firefox extension to provide user-to-user encryption to Dropbox, Facebook, and email. ShadowCrypt~\cite{he2014shadowcrypt}, MessageGuard~\cite{ruoti2015messageguard}, and Virtru~\cite{Virtru} are browser extension-based solutions that use an overlay to encrypt messages exchanged through any browser application. The issue with browser-based solutions is that (1) they do not work with non-browser-based applications, and (2) they inherit all browser-based attacks. There are many instances of malicious browser extensions used by millions of users on official Chrome/Firefox extension stores~\cite{ositcom,awakesecurity,catonetworks}.

\paragraph{Application independent encryption for phones}
Mimesis Aegis~\cite{lau2014mimesis} and Babelcrypt~\cite{ozcan2015babelcrypt} authors leverage accessibility services to create transparent overlays over all the applications, intercepting all communication and performing encryption/decryption before the data reaches the application. 
First, these are only explicitly designed for phones and cannot be directly incorporated into computers where there is a significant amount of user-to-user communications through apps such as Slack, Discord, Microsoft teams, and emails. 
Second, we believe that such unintended use of permissions to achieve security is not feasible for mass deployment~\cite{accessibilityservicesremove}. The other scalability issue is that it has to update the GUI for every application it supports and when there are major app updates to the GUI. Also, research is needed to determine if a malicious app can update the GUI to break the overlays. Finally, the system assumes a trusted client.

Our design assumes active untrusted client apps to secure all apps on the computer in an application-independent manner. Furthermore, it does not rely on any application-specific GUI, making it more scalable. Additionally, our design intentionally shows the encryption process to users to gain their trust in the system, which could improve adoption.


\paragraph{Usability of secure communication}

Large-scale system adoption depends on the ease of using the system~\cite{unger2015sok}. A complicated system that is hard for users to understand and use leads to user errors, significantly decreasing the security of a system~\cite{ruoti2013confused, whitten1999johnny}. While developing a system, it is important to design it so that users are willing to use it and less likely to make mistakes. Prior research shows this can be achieved through informative messages for first-time recipients of secure messages, intuitive interfaces, and integrated, content-sensitive tutorials~\cite{ruoti2015helping, ruoti2015johnny}

Research shows that users prefer a system that tightly integrates with existing web applications ~\cite{atwater2015leading, ruoti2015helping, ruoti2013confused, ruoti2015authentication}. Research also shows that most users are not interested in encrypting all of their online data~\cite{gaw2006secrecy, ruoti2013confused}. As such, users must be able to control when their data is encrypted. While designing an encryption system, the additional cognitive load (UI, UX) should be minimized. If encryption gets in the way of users completing tasks, it is more likely that they will not use it~\cite{herley2009so}.

\paragraph{Trusted I/O}
Trusted I/O paths such as Fidelius~\cite{eskandarian2019fidelius} ensure the confidentiality of data from an untrusted operating system. For example, if a user sends a message on Slack using Fidelius, the message is encrypted before it reaches the OS and is decrypted before it is sent to the Internet. So, the data is protected from a compromised OS and malware, but the Slack server sees everything in plaintext. Furthermore, it requires hardware changes on the keyboard and the display end. 
InfoGuard assumes a trusted OS and untrusted non-sudo applications. Fidelius or other trusted I/O path systems can be used along with InfoGuard to provide E2EE in an application-independent manner with an assumption of untrusted OS. 

\paragraph{Centralizing control of encryption and security parameters}


Previous research~\cite{fahl2013rethinking,conti2013mithys,o2017trustbase,bates2014securing} proposes centralizing user/administrator control of TLS certificate validation for all applications. This approach removes the burden on developers to implement correct validation and allows administrators to customize certificate validation by employing plugins that strengthen validation (e.g., revocation checks, DANE [13], etc.).
SSA~\cite{o2018secure} provides administrator control of additional aspects of TLS (version, ciphers, extensions, sessions, etc.) and can easily combine with centralized TLS validation systems like TrustBase~\cite{o2017trustbase}. SSA enables developers to increase the security of their applications by (1) allowing them to use a centralized and simplified TLS implementation and (2) modifying the TLS parameters set by an administrator, but a developer cannot decrease the security set by the administrator.

Similar to SSA~\cite{o2018secure}, we provide an E2EE API to help applications easily use E2EE for communication. Administrators/users can enable the E2EE without application support. Unlike TLS, encryption in our system works even if the server lacks support for encryption. In previous work like the SSA, users depend on the server to implement TLS before using encryption. Users do not need to trust the service provider to handle sensitive data properly. Our system offers users a single point of control over all the applications that can read their messages, enabling them to keep sensitive data private.

We evaluated 52 popular apps used for personal~\cite{personal_apps} and business~\cite{business_apps} communication. 
Only five apps provide E2EE by default (see Table~\ref{table:app_analysis} in the Appendix). 
Among personal apps, only three provide E2EE by default, five offer it as an optional feature, and the remaining 15 do not provide it. E2EE support is even less for business apps---two provide it by default, and one as optional. 

All analyzed applications, including those with E2EE, depend on trusting client applications. Users input plaintext in these apps, which they encrypt before sending on the Internet. This is undesirable as true E2EE should not rely on client app trust, akin to how we do not trust these applications's servers.

\section{Reddit Threads Analysis}


\paragraph{\textbf{Research question}}
Our first step was understanding the concerns, attitudes, and obstacles to adopting E2EE among privacy-conscious individuals.



\paragraph{\textbf{Methodology}}



We analyzed Reddit discussions about the privacy of online communication platforms from three prominent subreddits: \textit{privacy} (1.3M members), \textit{privacytoolsIO} (206k members), and \textit{privacyGuides} (55.3k members). Given that the participants voluntarily engaged in privacy-related discussions, we assumed this captures the perspectives of privacy-conscious individuals.

First, we gathered all posts from a three-year period (April 2020-2023). We selected relevant posts for analysis based on case-insensitive searches for common keywords related to secure communication, such as "e2ee" and "messaging." We also included keywords for various communication apps with different properties, such as "signal," "whatsapp," "telegram," "discord," and "microsoft teams," to gain insight into users' thoughts about these apps and their properties.
For phrases longer than one word, we checked if the words in the phrase occurred within a single sentence in a post. The mean number of comments per post was 13.86, and the mean score (the difference between upvotes and downvotes) was 43.66. However, 75\% of the posts had less than 13 comments and a 10 score. A link to the search scripts will be made available in the final version of the paper.

We used open coding to identify topics, followed by thematic analysis to identify users' concerns and factors hindering E2EE adoption. 
Two researchers independently coded all the posts and their comments and resolved differences through discussions. In total, we analyzed 210 posts with 98.5 comments each on average.
Our goal was not to draw generalizable conclusions about the prevalence of specific issues but to identify some factors that hinder E2EE adoption.

We initially selected a random sample of 100 posts for qualitative analysis and repeatedly added ten new posts until we reached saturation and no new codes were identified. We aimed to consider all posts regardless of their popularity. However, a post with multiple upvotes signifies multiple people's opinions. To account for this, we created a pool of posts from which we selected a random sample for analysis, with each post represented in the pool as many times as its score, which was the difference between upvotes and downvotes.


\paragraph{Results}
Our qualitative analysis revealed six concerns.

\paragraph{C1: Reputation and no open-source}

Many communication apps are available on the market, and many users believe none are secure. Therefore, most users choose apps using a process of elimination. Our analysis showed some users choose an app based on the following properties besides E2EE: (1) software is open source, (2) reputation of the company and CEO, (3) app's historical reputation, (4) location of the app's principal shareholders, (5) location of app's servers, and (6) encryption algorithm reputation. 

Closed-source apps claiming to be secure and implementing E2EE do not automatically garner users' trust. Since users cannot verify the app's privacy and security claims, some users stated that trusting these apps more than others is futile. Only four apps in Table~\ref{table:app_analysis} are open source (Signal, Telegram, Zulip, and Mattermost). Furthermore, E2EE is offered in one app by default and one app as an option. Most commercial apps do not release source code to maintain a competitive advantage. Open-sourcing the code could lose the advantage and risk legal disputes. Therefore, there is a need to isolate the encryption code from other proprietary code and make it open source. This approach could help increase users' trust in apps providing E2EE.

    \begin{quote}
        ``\textit{Even if an app such as Telegram does not implement E2EE by default is trusted more than WhatsApp because its open source.}''  
    \end{quote}
    
    \begin{quote}
        ``\textit{Telegram is better than Whatsapp. At least it’s open source.}''
    \end{quote}
    
    \begin{quote}
    ``\textit{you first have to trust them not to implement a backdoor in the encryption, closed source software + facebook behind it is the perfect recipe for not trusting it}''
    \end{quote}

When two apps have similar properties, users are more likely to distrust an app owned by someone with a bad reputation or a history of security breaches. 
    
    \begin{quote}
        ``\textit{The privacy issue with Whatsapp is, it's owner by Facebook which is exactly opposite of privacy and now whatsapp will share the data with it.}''
    \end{quote}

Users expressed concern regarding certain E2EE algorithms. For example, Telegram's proprietary algorithm was broken by security researchers, causing some to question its reliability and trustworthiness even after updates. Not everyone agrees on their preferred algorithm.


Users revise their opinions when things change. For instance, when Facebook bought WhatsApp, some trusted WhatsApp less. When Telegram moved its headquarters to Dubai, users doubted their intentions. 
    
    \begin{quote}
        ``\textit{Telegram is evil, not automatically E2E and based in dubai middle eastern governments will savagely hack those servers some day}''
    \end{quote}

    \begin{quote}
        ``\textit{Rakuten (the owners) do collect metadata. But then again, they are not Facebook and are not American, so that alone makes them a lot more trustable than WhatsApp/FB Messenger.}''
    \end{quote}

\paragraph{C2: E2EE clients have plaintext access}

Many users expressed concern that client apps, including the state-of-the-art Signal libraries, have plaintext access. Given this access, they do not perceive value in using E2EE over non-E2EE apps. In addition, some users argue that E2EE provides no additional privacy benefits, so limiting themselves to a select few E2EE apps lacking features and performance is unjustified.

    \begin{quote}
         ``\textit{I think the bigger threat from companies claiming to offer "end-to-end" encryption is that the company actually controls the "ends", not you, and so they can access your data unencrypted if they really want to.}''
    \end{quote}
    
    \begin{quote}
        ``\textit{It depends on who has access to the decryption keys. FB has the decryption keys, so the E2E encryption is not good enough, if you want full control.}''
    \end{quote}

\paragraph{C3: E2EE is invisible}
Some users lamented their inability to distinguish between an E2EE and a non-E2EE app. They cannot discern what new security features these apps offer that were not available in previous apps that suffered from data leaks and breaches. Non-expert users who lack extensive experience with privacy or security technologies are especially hesitant to trust E2EE. It is difficult for them to accept that some apps are more secure than others. A reason for this lack of trust is that there is no visible difference in privacy between apps that offer E2EE and those that do not. Some users mentioned that even if they could muster the confidence to trust the app based on experts' opinions, it is arduous for them to persuade their family and friends as there is no feasible method to demonstrate the relative security superiority of one app over another. Thus, if they cannot convince their contacts to adopt E2EE, they have no reason to use it.

    \begin{quote}
        ``\textit{No one really knows if the app really provide E2EE}''
    
    \end{quote}

\paragraph{C4: E2EE adoption overhead}
Many users wanted to use privacy-enhancing apps to share their data online. However, the high cost of switching to these apps often prevented them from doing so. Three primary costs were identified:
(1) The burden placed on their contacts since all of their friends and family were using an insecure app and were unwilling to switch to a different one due to convenience, preferred features, or a lack of trust in less popular apps.
(2) The loss of features, as switching to a secure app meant giving up important features, such as multiple profiles, that were convenient or essential for business use.
(3) The learning curve, as many (especially non-technical) users find it challenging to adapt to new apps.
Users frequently voiced concerns that switching apps solely for privacy or security is impractical.


    \begin{quote}
        ``\textit{Just wish Signal supported multiple profiles, like Telegram. It would be great for dual-sim devices.}''
    \end{quote}

    \begin{quote}
            ``\textit{Discord is going to be like YouTube for me. I'll still use it because there's no good alternatives.}''
    \end{quote}

    \begin{quote}
        ``\textit{Literally none of my friends want to switch to matrix, I tried convincing them but they refuse, they don't care about privacy and gladly trade it for the features that discord allows. Unfortunately there is no current alternative that is better or equal to discord, otherwise people would switch.}''
    \end{quote}

\paragraph{C5: No access to E2EE apps}
Privacy-focused communication apps are most needed in countries with high levels of censorship. However, these countries ban many of these apps, forcing users to rely on less secure communication options.
    
        \begin{quote}
            ``\textit{What is the use. My country will simply ban the app.}''
    \end{quote}

Some individuals were constrained to use an app they felt lacked privacy due to organizational mandates. Some users challenged the organization but were unsuccessful. Application-independent privacy frameworks could address such issues by allowing institutions to select applications that align with their requirements while enabling users to make security choices that meet their needs.

    \begin{quote}
        
       ``\textit{I ditched whatsapp completely almost 2 months ago but now my college asks me to use whatsapp for reasons which I can't ignore such as for assignments, schedules and class groups, etc. I tried to persuade them to use Signal instead but it didn't go very well.}''
    \end{quote}

\paragraph{C6: E2EE lacks coverage}
Many users are interested in more data privacy in apps other than instant messaging. To achieve this, they must find an E2EE app for each use case, such as writing notes and syncing contacts. Furthermore, some apps lack desired features, such as Proton Mail not encrypting the email subject line. Finding and verifying different apps for different use cases is a burden. 

    \begin{quote}
        ``\textit{Am I crazy to be worried about my privacy when it comes to note taking apps?}''
    \end{quote}

\section{\sysname}

This section describes the system and adversary model, design goals, and system design for \sysname, a system supporting user-controlled, application-independent encryption.

\subsection{System and Adversary Model}

Our system model is a client-server architecture in which two clients communicate through a centralized server. The clients are software applications installed on user devices, like Slack, Outlook, or webpages accessed through a web browser. The servers act as intermediaries to facilitate client-to-client communication, such as messaging and email.

For client-to-client communication, the user enters the message using an input device such as a keyboard. An input device driver, a software component in the operating system (OS), enables communication between the OS and the input devices by translating signals the input device sends into a format the OS understands.

The OS then passes the message to the windowing manager (e.g.,~X11, Wayland) that passes the plaintext input to the corresponding application. The application then encrypts the message for transmission over TLS and sends it to the application server. The server decrypts the message, re-encrypts it for the TLS connection to the recipient, and delivers it to the recipient. Upon receiving the message, the recipient's application decrypts it, allowing the recipient to read it.

Some applications (e.g.~Signal) support End-to-End Encryption (E2EE). In these applications, the message is encrypted with a key shared only between the sender and recipient before it is encrypted for transmission over TLS. Thus, the server cannot access the plaintext user data after decrypting the TLS transmission.

\paragraph{Adversary model}
Adversary \A is an active global attacker that controls an application server with plaintext access to all messages and metadata flowing through it. \A also controls client applications having user (non-privileged) access.

The adversary could be (1) law enforcement or an oppressive regime coercing an application or (2) hackers compromising an application. 
The adversary's goal is to compromise message confidentiality and integrity to be able to read and modify users' messages.

\subsection{Design Goals}
The following design goals are based on our qualitative analysis of Reddit threads and prior research.

\paragraph{G1: Encryption isolation and no plaintext access by client app}


We address concerns C1 and C2 by isolating encryption in a single security module that first receives sensitive user input from the OS. The security module delivers only encrypted text to the client app. The isolation allows (1) apps to open-source only the security module for auditing purposes and to gain user trust, (2) security experts can develop the security module, (3) smaller codebase for auditing, and (4) simpler and fewer security patches. Users and OS developers can choose a security module they trust.

Most E2EE apps, like Signal, support client-side encryption to prevent the server from having plaintext access. If the server is not trusted, it begs the question of why we trust the client app with plaintext access.
Therefore, designing an app that prevents client-side plaintext access reduces the risk of a compromised client.

\paragraph{G2: No mandatory app support and per-app configuration}
Previous research~\cite{abu2017obstacles} shows that usability is not the primary adoption obstacle; fragmented user bases and lack of interoperability are significant obstacles. Our Reddit analysis found similar concerns. Apps that users need do not provide E2EE (C4, C5, and C6).

\sysname aims to provide application-independent encryption, which means that it should not depend on the specific details of an application, such as its GUI, and should not require an application to cooperate and provide encryption. This property allows \sysname to scale to all applications and achieve E2EE even if an application does not provide it. We will use low-level keyboard hooks to intercept keyboard events at the system level and perform encryption and decryption operations to achieve this goal. This approach allows \sysname to work independently of any specific application. It gives users complete control over encryption, such as what algorithm to use, what to encrypt, and when to encrypt.


\paragraph{G3: Visible encryption}
Many users may need help understanding how E2EE encryption works. In our Reddit analysis, we noticed users' concern (C3) regarding their inability to distinguish between E2EE and non-E2EE communication. A quantitative study by Abu-Salma \etal~\cite{abu2018exploring} found that three-quarters of respondents believed unauthorized entities could access their end-to-end encrypted communications.
A study by Dechand \etal~\cite{dechand2019encryption} of E2EE on WhatsApp shows that users do not trust encryption as currently offered. We propose making encryption more visible to increase awareness and gain users' trust. Ruoti \etal~\cite{ruoti2013confused} found that users tend to trust manual encryption more than invisible encryption because they can see how it works, stating that ``\textit{Surprisingly users were accepting of the extra steps of cutting and pasting ciphertext themselves. They avoided mistakes and had more trust in the system with manual encryption. Our results suggest that designers may want to reconsider manual encryption as a way to reduce transparency and foster greater trust.}'' The critical difference between manual and invisible encryption is that manual encryption allows users to see the encrypted text, ensuring that encryption occurs. Users only pass encrypted text to the application, ensuring it never sees their plaintext data. We observed similar results in our Reddit analysis, where users could not distinguish any differences in the security and privacy offered by E2EE apps due to invisible encryption.

Atwater \etal~\cite{atwater2015leading} conducted a study that revealed that most users (69\%) did not prefer trust between standalone manual encryption and integrated solutions. However, among the remaining 31\% of participants who had a preference, the majority (10 out of 11) trusted the standalone manual solution more. These findings suggest that while visible encryption may not be necessary for all users to gain their trust, a significant proportion of users require a visible encryption system.

On the other hand, users often reject security protocols due to the perceived effort required outweighing the extra security gained~\cite{herley2009so}. To address these concerns, \sysname aims to minimize user effort by automating the encryption/decryption process while promoting user trust by providing information about the encryption algorithms and key management techniques. Additionally, \sysname shows encrypted text to the user in the corresponding application's textbox, further promoting translucency and trust in the system.

\paragraph{G4: Desktop app}
\sysname aims to increase user adoption and usage by leveraging user trust in desktop apps, as research has shown that users perceive desktop applications to be less likely to transmit their sensitive messages back to the developer of the software~\cite{atwater2015leading}. 
This perception of trust in desktop apps may be due to their offline nature and the fact that users have more control over the software running on their computer~\cite{atwater2015leading}. 
Therefore, we designed our protocol as a desktop app to take advantage of this trust and to increase the likelihood of users adopting and using our system.

\paragraph{G5: Selective encryption}
Users may be interested in encrypting only some of their online data~\cite{gaw2006secrecy}. 
Therefore, \sysname aims to provide users with easy-to-use controls over when their data is encrypted, allowing them to encrypt specific conversations, messages, or even part of a message selectively. 
Fine-grained selective encryption means that only the sensitive information must be encrypted. Applications/servers can still use the non-sensitive information to provide a better user experience. Furthermore, it gives servers and monitors more context to trust legitimate encryption uses.

\subsection{System Design}

To achieve our design goals, we propose \sysname - an input driver extension that ensures that all the data generated from that input device is secured and private from even the client apps. The design and implementation in this paper centers on keyboard input; however, the ideas can be readily adapted to other input devices. 

We explored two approaches with the same security properties but a different UX: (1) \sysname{v1} -- dual GUI display, user types in the application GUI with plaintext displayed in $\sysg$ and ciphertext updated and displayed in $GUI_{app}$ as it is entered. (stream cipher) (2) \sysname{v2} -- single GUI display where $\sysg$ is shown on top of the screen where the user enters plaintext, and when finished, the ciphertext is placed in the $GUI_{app}$ (block cipher). 

\subsubsection{\sysname}

This section describes the high-level design of \sysname. 
It has two components: Interceptor and $\sysg$.
Fig~\ref{fig:system} in the Appendix shows all the entities of \sysname and the message flow. 

\paragraph{\textbf{Interceptor}}
    The interceptor performs the following tasks:
    \begin{enumerate}
        \item \textit{Listens for encryption/decryption shortcuts}: It passively listens to the read-only keyboard input file to detect when an encryption or decryption shortcut is pressed. 
        \item \textit{Input interception}: After encryption mode is initiated, the Interceptor gains \textit{exclusive} access to the keyboard input buffer. The buffer allows applications to receive raw keyboard events, including key presses and releases, without any processing or interpretation by the system. Generally, windowing systems such as X or Wayland read keyboard input files, and applications receive keyboard input events from these windowing systems. The Interceptor needs to ensure that neither the windowing manager nor any other non-sudo application should be able to access the keyboard input from the keyboard input file. 
        \item \textit{Plaintext transmission to $\sysg$}: The user first selects the recipient who will receive the message. 
        Interceptor must ensure that the transmission from Interceptor to $\sysg$ has confidentiality and integrity properties from any non-sudo application on the client machine, including the target app where the user encrypts the input.\\
        \item \textit{Ciphertext transmission to windowing system}:
        The Interceptor transfers ciphertext to the application through the windowing system.
        While passing the encrypted text and metadata to the target app, Interceptor wraps it in the "Guard-start<encrypted text>Guard-end" format to allow the sender and the recipient to identify the encrypted text uniquely.
        Interceptor maintains a list of permitted keys, such as arrow keys or shortcuts Ctrl+A, and Ctrl+C, that it does not encrypt. These keystrokes are passed to apps through the windowing system in plaintext. 
        \item \textit{Decryption}: When a user presses the decryption shortcut, the Interceptor retrieves the user-selected text, decrypts it, and transmits the plaintext to $\sysg$. 
    \end{enumerate}

\paragraph{\textbf{$\sysg$}}
\begin{enumerate}
    \item \textit{Displays plaintext}: $\sysg$ receives the plaintext from the Interceptor during encryption and decryption. It displays the plaintext to the user character by character during encryption and all at once after decryption. $\sysg$ ensures that other non-sudo applications cannot access plaintext through memory or screen recording. 
    \item \textit{Set parameters}: Allows users to set different parameters such as recipient, encryption algorithm to use, and key size. Users must choose the recipient every time they encrypt or decrypt. 
    $\sysg$ is responsible for getting the recipient's identity after the encryption is initiated and passing it to the Interceptor (see Fig~\ref{fig:infoGuard_v1_0} in Appendix). 
    The rest of the parameters can be set for the long term or recipient-based and can be reused without selecting them every time.
\end{enumerate}

A malicious app cannot retrieve plaintext through the windowing manager by listening to keystrokes. Even if a malicious app creates a malicious overlay on top of $\sysg$, it cannot access plaintext because whenever the user presses a key the Interceptor intercepts it and passes it to the authenticated $\sysg$ only.

We designed two prototypes to experiment with different interfaces for $\sysg$ and how the interceptor sends ciphertext to the application. The two prototypes are {\sysname{v1}} and {\sysname{v2}}.

\subsubsection{\sysname{v1}---Continual Encryption}
Fig~\ref{fig:infoGuard_v1} in Appendix shows the screenshot of \sysname{v1}. 
When the user types their plaintext message, Interceptor forwards it actively to the $\sysg$, which displays the plaintext characters that users type in the target application. 

One of our design goals is to gain users' trust through translucency, where we visually show users that the target application only receives the encrypted text. The Interceptor performs character-by-character encryption as the user types and sends the encrypted characters and metadata dynamically to the target application to give real-time feedback to users as they type (Fig~\ref{fig:infoGuard_v1}).

\begin{figure*}
\centering
\includegraphics[width=0.62\textwidth]{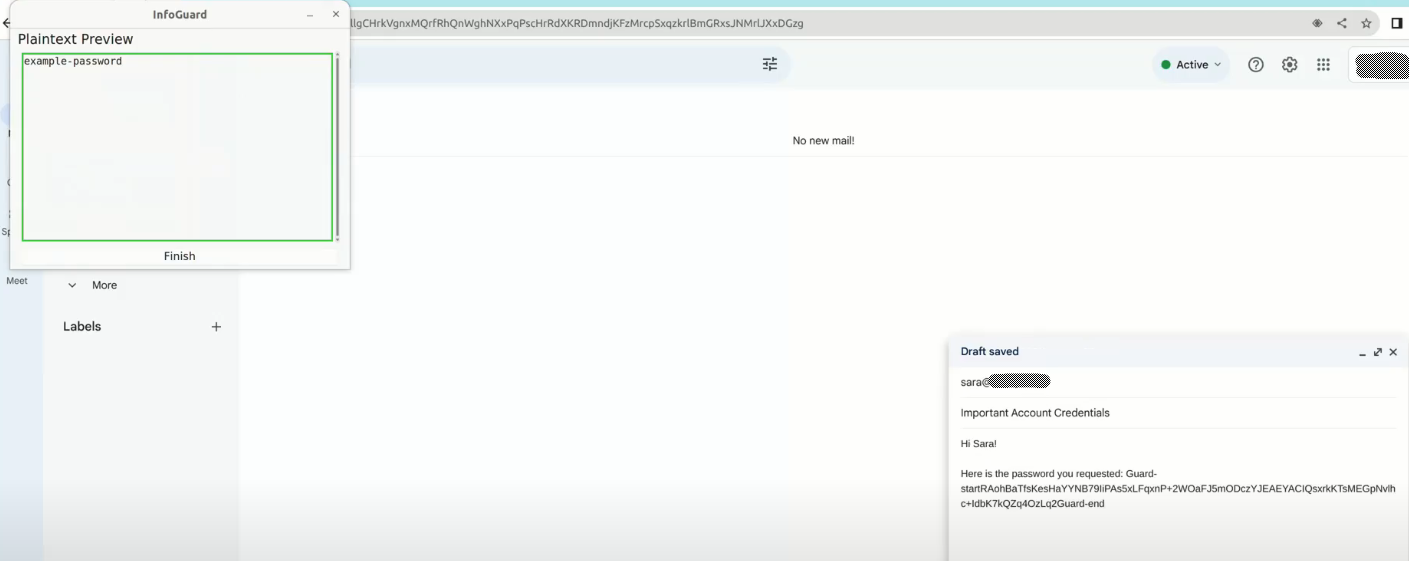}
\caption{\sysname{v1} encryption. The user has already selected the recipient on the previous screen of $\sysg$.}
\label{fig:infoGuard_v1}
\end{figure*}

\begin{figure}
\centering
\includegraphics[width=0.48\textwidth,trim={1cm 2cm 4cm 0},clip]{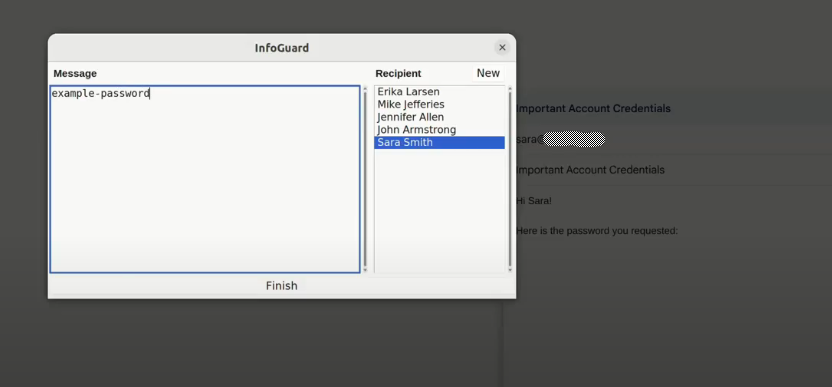}
\caption{\sysname{v2} encryption.}
\label{fig:infoGuard_v2}
\end{figure}

\subsubsection{\sysname{v2}---One-time Encryption}
Fig~\ref{fig:infoGuard_v2} shows the screenshot of \sysname{v2}.
\sysname{v2} has all same entities as in \sysname{v1}, however the input and flow of messages are different. In \sysname{v1}, the plaintext messages are typed on the target application textbox but the Interceptor passes the encrypted text to the target app and plaintext to $\sysg$. However, in \sysname{v2}, Interceptor allows the user to type in a textbox in $\sysg$, making it easier to edit the plaintext,  and encryption takes place only after they finish typing. 

When a user initiates encryption mode, $\sysg$ appears as a visible overlay on top of the entire screen. A user cannot interact with any other GUI until they close the $\sysg$. This window has an option to select the recipient and a textbox where the user enters the plain text they want to encrypt. In this version, the interceptor does not perform stream cipher, instead, it intercepts and forwards all plaintext to $\sysg$ and performs block cipher encryption when $\sysg$ notifies it after the user presses the finish button. The Interceptor encrypts the message using the signal protocol and passes the base64 encoded encrypted text and metadata to the target application's text using the virtual keyboard automatically.


\subsubsection{Discussion}

\paragraph{Key management}
\sysname employs the Signal protocol for encryption across multiple applications, allowing two users to communicate securely over different apps. In terms of key management, \sysname can use two options. Firstly, \sysname can generate a new public-private key pair for each user, as is done in the Signal protocol, and the sessions are updated across the applications. Alternatively, the client can maintain a different public-private key pair for every user for every app. However, the second approach offers no additional security benefits, as all the keys are stored in the same location on the client's machine, which can be a single point of failure.

On the other hand, the first approach does not pose any significant privacy risks compared to the second approach. In the second approach, every client uploads a different public key for every app they use, which allows others to retrieve those keys to check if a user uses that app. This could potentially reveal sensitive information to unauthorized parties. Therefore, we include the first approach of generating a new public-private key pair for each user in \sysname.

Moreover, utilizing the same key bundle across different applications, \sysname simplifies the key management process and reduces the system's complexity. A client in \sysname only needs to securely store the same key bundle as the Signal protocol.

When a pair of users switch their communication from one app to another while using \sysname, the same session is used, and keys keep ratcheting as in the Signal protocol. This means the keys are updated and advanced every time a message is sent or received, ensuring the messages are encrypted with a fresh set of keys. This mechanism provides forward secrecy in the Signal protocol.

We use a centralized key server to distribute public keys. A fake key distribution attack can be detected or prevented using secure key distribution techniques such as CONIKS~\cite{melara2015coniks}, SEEMLESS~\cite{chaseseemless}, or KTACA~\cite{yadav2022automatic}. 


\paragraph{Windowing manager-based attacks:} 
New window managers, such as Wayland, prevent manager-based attacks by preventing third-party applications from taking screenshots or recording screens. Wayland has been the default window manager in Linux OSes since Ubuntu 21.04, Fedora version 25, Debian v10, and Red Hat Enterprise Linux v 10. Windows and macOS operating systems already restrict an app from accessing another app's GUI.

However, privacy risks exist for some older window managers, such as X11, due to keylogging and screen capturing. As the X window manager is a shared resource, all clients have equal access, meaning any application can access $\sysg$ graphically through screenshots or screen recordings. Currently, there is no way to prevent this. 
To prevent X-based attacks in old windowing managers requires the $\sysg$ process to pass plaintext directly to the display frame buffer, bypassing the X server. This approach prevents any application from scraping text from the display or capturing screenshots and recordings, as the X server has no access to the plaintext. However, we did not implement this in our prototype.

\section{Implementation and Performance Analysis}

We built proof-of-concept prototypes of both versions of \sysname to demonstrate their feasibility, evaluate their performance, and conduct user studies. Appendix~\ref{app:implementation_mess_flow} describes messages flow and Appendix~\ref{app:implementation_details} describes in-depth implementation details of \sysname{v1\& v2}.

Our implementation of the Interceptor is a native C/C++ application that runs the $\sysg$. We employed the Tkinter Python library for Tcl/Tk to develop a user-friendly and adaptable design with cross-platform compatibility for $\sysg$. We utilized the Signal C library (libsignal-protocol-c)~\cite{signalclibrary} to implement forward secrecy encryption and decryption, which we modified to support a stream cipher In \sysname{v1}. 




To determine their robustness, we tested our prototypes across the following communication apps and websites. Our prototypes work on Gmail, Slack, Facebook, Microsoft Teams, and Discord. We also successfully tested our prototypes on Google Docs, where a user encrypts the sensitive part of a document before sharing it with another person. Only the recipient can decrypt the sensitive data.


\textit{Encryption Latency:} \sysname introduces three negligible delays. \sysname continuously monitors the keyboard input file for encryption or decryption shortcuts. This process adds only 139 microseconds to each keystroke, which is imperceptible to the user. In \sysname{v1}, every keystroke is encrypted in real-time, adding only 107 additional microseconds to XOR for each character with the precomputed key stream. We ran the measurements 1,000 times and reported the mean value.

For \sysname{v2}, the Interceptor performs a block cipher on the complete plaintext. In our analysis, it took 137 and 140 milliseconds to encrypt 200 and 1000 characters of plaintext, respectively. To decrypt 50 and 1000 characters of plaintext in \sysname, the Interceptor took 707 and 735 milliseconds, respectively.



\textit{Storage Overhead:} \sysname introduces two types of storage overhead: (1) users' devices need to store decrypted data, and (2) the application servers need to store additional \sysname metadata transmitted by the user. The extra metadata being transmitted is equivalent to the additional network data that the user will send or receive while transmitting the data encrypted through \sysname.

The space used to store decrypted text on the user's device will depend on how much data they encrypt using \sysname. Essentially, this overhead is proportional to the size of the data. Additionally, users can store data backups with symmetric encryption on other cloud storage services.

\textit{Network Overhead:}
While \sysname maintains a one-to-one mapping from plaintext to ciphertext, it converts the encrypted text to base64, which maps 3 bytes of data to 4 bytes while encoding, essentially adding 33\% overhead.
Furthermore, the Signal protocol's metadata contains the following components: message version, ciphertext current version, ratchet key, ratchet key length, counter key, counter value, previous counter key, previous counter value, ciphertext key, ciphertext length, and MAC. The approximate total size of the metadata is $50 + \lfloor{\frac{\text{plaintext length}}{128}}\rfloor + 2\lfloor{\frac{\text{message number}}{128}}\rfloor$. The total size of network overhead is $0.33\times \text{plaintext length}  + 50 + \lfloor{\frac{\text{plaintext length}}{128}}\rfloor + 2\lfloor{\frac{\text{message number}}{128}}\rfloor$

\paragraph{\textbf{Limitations}}

One of the limitations of \sysname is that, because it is application-independent, it lacks awareness of the contexts in which users initiate encryption. This lack of context can result in unexpected behaviors or errors when a user encrypts parts of a message. For example, the application may throw an error if a user encrypts a field that expects a specific length digit input, such as a zip code. Another example is a user encrypting an email address in the \textit{TO} field while sending an email. An in-depth user study is needed in the future to analyze users' behaviors when they use \sysname across different websites and how \sysname can be improved to inform users where they can or cannot use \sysname. Enabling user-controlled encryption raises many research questions to address to make it practical.

Another limitation is that organizations often need to scan content to control the release of sensitive information. They may prohibit the use of InfoGuard for this reason.

\paragraph{Apps where \sysname could be useful}
We evaluated 52 popular apps used for personal~\cite{personal_apps} and business~\cite{business_apps} communication. 
Only five apps provide E2EE by default (see Table~\ref{table:app_analysis} in Appendix). 
Among personal apps, only three provide E2EE by default, five offer it as optional, and the remaining 15 do not provide it. Two business apps provide it by default, and one as optional. 

All of these apps trust the client with plaintext. Users can use \sysname to selectively encrypt sensitive content on any of these apps to increase privacy while still using existing features.

\section{Evaluation}
This section compares \sysname with existing E2EE systems based on predefined security, scalability, and perceived usability metrics based on our design goals. The comparison is presented in Table~\ref{table:system_comparison}, but it is not intended to be exhaustive. Based on our analysis of Reddit user feedback, we focus on the properties that we deem essential for promoting the widespread adoption of E2EE among privacy-concerned users. The existing systems we consider are MessageGuard, Shadowcrypt, Mimesis Aegis, and Babelcrypt. Because they were rated similarly for our metrics, we refer to them as "Existing application-independent E2EE systems".

\newcommand*{\yes}{\colorbox{green!20}{\tiny\CIRCLE}}
\newcommand*{\no}{\colorbox{red!20}{\tiny\Circle}}
\newcommand{\sometimes}{\colorbox{gray!20}{\tiny\LEFTcircle}}

\begin{table}[tbp]
    \caption{Comparison of \sysname with existing E2EE systems that are also app-independent. App-based standard E2EE protocols are included for a baseline comparison.}
    \begin{center}
    \resizebox{0.48\textwidth}{!}{%
    \begin{tabular}{|c |c |c |c|}
        \hline
        
        \textbf{Properties} & &\multicolumn{2}{c|}{\textbf{App }}\\
        & &\multicolumn{2}{c|}{\textbf{Independent}}\\

        
        &  \rotatebox[origin=c]{90}{\textbf{App based}} & \rotatebox[origin=c]{90}{\textbf{Existing}} & \rotatebox[origin=c]{90}{\textbf{\sysname}} \\
        \hline
        \textbf{Security \& Privacy} & & & \\
        \hline
        Prevents server access & \yes & \yes & \yes  \\
        Prevents client app access & \no & \sometimes & \yes\\
        User-controlled app-independent encryption & \no & \yes & \yes \\
        Easy security evaluation and patching & \sometimes & \yes & \yes \\
        \hline
        \textbf{Scalability} & & &  \\
        \hline
        Requires no per-app configuration & \no & \no & \yes \\
        \hline
        \textbf{UX} & & &  \\
        \hline
        Supports selective encryption & \no & \sometimes & \yes \\ 
        Supports visible isolated encryption &  & \no & \yes \\
        Preserves target application UX & \yes & \sometimes & \no \\
        \hline
    \end{tabular}
    }
     \yes = support, \sometimes = limited support, \no = no support 
    \end{center}
    \vspace{-0.5cm}
    \label{table:system_comparison}
\end{table}

\subsection{Security \& Privacy}
\textit{Server access:} Standard E2EE protocols are designed to prevent data access from the servers. App-independent E2EE systems, including \sysname, build on these protocols to achieve the same property.

\textit{Client access:}
An application can execute two types of attacks on the client side: passive and active. Passive attacks involve reading user text while they type in the app and sending it to the server or having bugs that allow other attackers to access plain text passively. Active attacks involve the client app actively attempting to access the plaintext, even when the user does not enter it directly into the app. 
None of the standard app-based E2EE protocols prevent these attacks. Existing E2EE application-independent apps prevent a passive attack as they do not let the plaintext directly pass to the application. However, they could not protect against active attacks because they use an overlay on top of the app to intercept and encrypt plaintext. As a result, a malicious app can either read the plaintext by monitoring keystrokes in the windowing manager or creating an overlay on top of the E2EE system's overlay. 
\sysname, on the other hand, prevents plaintext access even against active attacks by intercepting and encrypting key presses before passing them to the windowing manager. The plaintext is passed directly to $\sysg$, preventing any other overlay from accessing plaintext.


\textit{User controls encryption:}
In app-based E2EE protocols, apps determine which data should be encrypted. In all application-independent systems, users decide what and when to encrypt and what algorithm to use.
Application-independent systems allow users to achieve security \& privacy on any application they want. Application independence refers to a system's ability to provide encryption without any action required from an application. This functionality makes it easier for users to adopt encryption without waiting for applications to update their security measures.

\textit{Easy security evaluation and patching:}
Centralizing encryption makes it easier to evaluate code and patch quickly if any vulnerability is found. Most app-based protocols, such as PGP, Signal, and OTR, have a small set of libraries that other applications can use. Analyzing and patching a few libraries fixes many applications, but each app must be updated (denoted by a half circle in the table). For application-independent systems, once the library is patched, the operating system on each device must be updated (denoted by a full circle in the table). 


\subsection{Scalability}
\textit{Requires no per-app configuration:} Existing E2EE systems require overlay designs that mimic the original app's GUI for every application they want to support. Additionally, they must update the overlay interface as frequently as the original app updates its GUI. Nonetheless, a malicious app can use frequent GUI updates to interfere with these systems. In contrast, \sysname works out-of-the-box with any application without requiring the app's GUI mapping.

\subsection{Perceived Usability}

The usability of a system is a critical factor in its adoption and value in the real world. Our analysis of perceived usability includes metrics that go beyond just the user interface (UI) to consider other important properties, such as user trust. 

\textit{Selective encryption:} In most app-based E2EE protocols, users cannot choose what part to encrypt; instead, they either encrypt everything by default or nothing. However, some apps (e.g.\ Telegram) allow users to choose what messages to encrypt. Existing application-independent E2EE systems also let the users decide what to encrypt, but they do not provide fine-grained control, such as encrypting part of an email. In \sysname, users can selectively encrypt even part of a message. For example, they can encrypt either the entire email or just one sentence in an email. While phone-based and browser-based systems also allow selective encryption to some extent, they do not provide the fine-grained control that \sysname does, such as encrypting part of an email.

\textit{Visible isolated encryption:}
App-based E2EE systems do not isolate encryption. Thus, isolation does not apply to them. Existing application-independent E2EE systems provide some form of visible encryption to improve the user experience. Previous research shows that users do not trust a system if everything happens in the background without visibility. In \sysname, users can see that plaintext is encrypted and only ciphertext is passed to the application. While complete visibility may not preserve the original app's UX, the UI remains the same. 

\textit{Preserve target application UX }
System designers choose between visible encryption or preserving the original app's UX. Application-based systems provide invisible encryption, while application-independent E2EE systems show users a colored dot, different background, or surrounding box to indicate that encryption is happening. However, users still type plaintext in the app and do not see encryption happening in real time. \sysname allows users to selectively encrypt their plaintext while offering complete visibility to increase users' trust and confidence in the system.

\section{User Study}
We conducted an IRB-approved laboratory study to assess the usability of both \sysname{v1} and \sysname{v2}. Additionally, we measured user trust and willingness to adopt. 

\subsection{Methodology}



We developed a desktop application using the Electron framework for our user study. This application encompassed demographic surveys, task instructions, and associated questionnaires. The application dynamically switched between different variants of \sysname based on the task. Ordering bias was mitigated by randomizing the sequence of Task 1 and Task 2.

Participants were asked to complete nine subtasks grouped into three main tasks:

\textit{Task 1:} Participants used \sysname{v1} to encrypt a message on three platforms -- Gmail, Discord, and Google Docs. Each subtask had a different scenario: (1) sending a Social Security Number (SSN) to a friend via Gmail, (2) transmitting a server password on Discord, and (3) adding a server password to a Google Docs document.

\textit{Task 2:} Similar to Task 1, participants used \sysname{v2} to encrypt a message on the same three platforms.

\textit{Task 3:} Participants used \sysname to decrypt a message on three platforms -- Gmail, Discord, and Google Docs.

At the beginning of the study, participants were informed about the nine sub-tasks. Participants were directed to a Desktop app providing detailed instructions for each task as they progressed through the study. The app automatically detected the current task and executed the corresponding \sysname version. We used test accounts for Gmail, Google Docs, and Discord. The automated setup minimized interaction with the study coordinator, increasing participants' privacy during the study. We believed more autonomy would produce realistic behavior. For each of the three tasks, we showed participants an instructional video ( ~40 sec) that demonstrated the usage of \sysname on Gmail.


\textbf{Recruitment}
 We recruited 24 participants by posting flyers on campus announcement boards across various departments. To facilitate communication, we specifically selected participants who were fluent in English. We excluded one participant from our study's dataset due to challenges in comprehending instructions due to limited English proficiency. On average, participants took 29 minutes to complete the study. We compensated each participant with a \$15 Amazon gift card.
 
\textbf{Demographics}
Out of 23 participants, 16 were 18-24 in age, while 7  were 25-34. Twelve participants identified as male and 10 as female. Additionally, 15 participants were undergraduate students and 8 were graduate students.

All the participants mentioned that they use some communication platforms such as Gmail or Discord every day, and keeping their messages private from these applications is important to them. Nine participants share sensitive information over these applications daily, 4 participants weekly, 2 participants monthly, and 8 participants rarely share sensitive information. Fifteen participants believed that communication apps could access sensitive data, 5 were neutral, and 3 believed apps could not. 

\textbf{Data collection and analysis}
During each task, participants were questioned about their comprehension of encountered errors and their approach to resolving them. In addition, their screen activity was recorded to facilitate later analysis. The collected free responses were analyzed using inductive coding and content analysis techniques.

\textbf{Limitation}
This user study exclusively focuses on assessing the usability of \sysname in everyday scenarios. Tasks that are infrequent, such as the installation of \sysname and the addition of new contacts, have not been included.

\subsection{Quantitative Results}

\textbf{Usability}
The median SUS score for the encryption using \sysname{v1} was 77.5, for the encryption using \sysname{v2} was 90, and for decryption was 95. 
We compared the SUS scores for the encryption process between \sysname{v1} and \sysname{v2} using a paired-sample t-test. There was a statistically significant difference between the SUS scores for \sysname{v2} encryption (M=87.5, SD=11.36) and \sysname{v1} encryption (M= 79.23, SD= 14.87); t(22) = 2.8, p=0.01. The mean difference in SUS scores was 8.26, and the 95 percent confidence interval was (2.16, 14.36).

\textbf{Trust}
Out of 23 participants, 21 were satisfied with the security and Privacy \sysname InfoGuard provides, with 17 being extremely satisfied. Furthermore, all participants agreed that using \sysname makes their communication more private. Their trust in both \sysname{v1} and \sysname{v2} was the same, with 7 participants somewhat agreed and 14 participants strongly agreed.


\textbf{Adoption}
Out of 23 participants, 17 said they were likely to adopt \sysname, whereas 4 said they were unlikely to adopt it. The remaining 2 were neutral.

\begin{figure}
\centering
\includegraphics[width=0.48\textwidth]{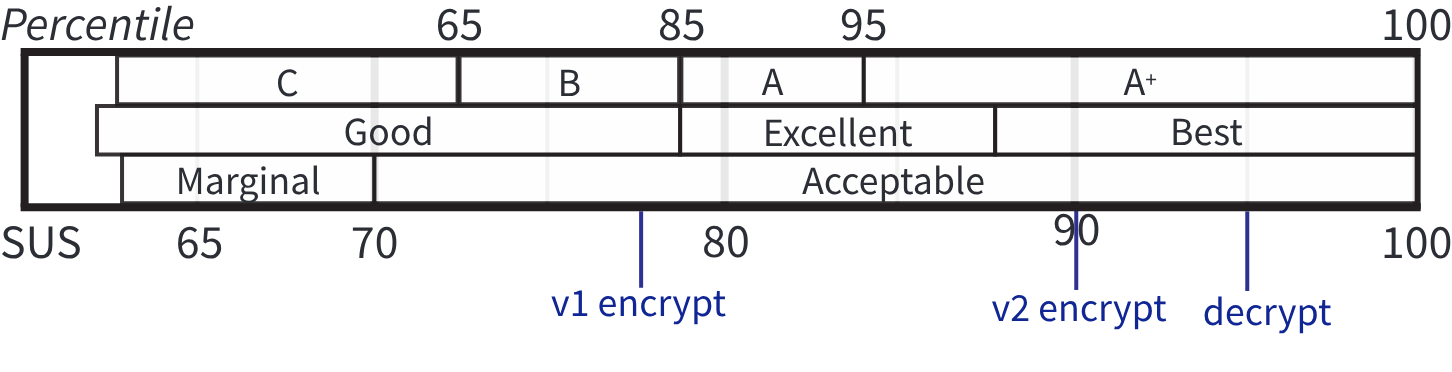}
\caption{SUS scores for \sysname{v1 \& v2} encryption and \sysname decryption.}
\label{fig:sus}
\end{figure}

\subsection{Qualitative Results}
We asked participants what they liked, disliked, and what could be improved. Overall, people liked the concept of \sysname and were willing to use and recommend it to their friends. 

\begin{quote}
    ``\textit{I really like the concept. It's strong suit is that it is extremely simple and easy to use. I would totally use this and recommend it to others if it was available.}''
\end{quote}

\begin{quote}
    ``\textit{I think this is a very cool idea and I would love to see it out on the market.}''
\end{quote}

Participants liked the simplicity and integration of InfoGuard encryption and decryption in the system. Participants explicitly mentioned that it was simple, fast, easy to use, and required minimal effort and learning curve. 

Participants liked the visible encryption, which was one of our main goals, and it increased their confidence and trust in the \sysname.

\begin{quote}
``\textit{I also really like how you can see the plaintext typed in the InfoGuard window while the encrypted text is shown in the messaging application, which feels like proof your message is going to be secure}''
\end{quote}

Participants also liked the selective encryption aspect of \sysname.

\begin{quote}
``\textit{It is a simple, straightforward to use interface and it allows you to encrypt data right in with the rest of your message, which is very convenient when using various platforms.
}''    
\end{quote}

Furthermore, some participants noticed and liked the application-independent nature of \sysname without us telling them. 

\begin{quote}
    ``\textit{I like that it doesn't depend on what kind of software I'm using. Even if it's a new messaging app that didn't exist when infoguard was built, I can still use it.}''
\end{quote}

Participants had a mixed reaction to using keyboard shortcuts to initiate encryption and decryption mode. Some participants were very positive about using shortcuts and believe it provides better integration. However, some participants who anticipated rarely using them prefer having buttons for encryption and decryption on the right-click menu. 

\begin{quote}
``\textit{It encrypts things! And only uses a shortcut key, which is nice}''
\end{quote}

\begin{quote}
    ``\textit{I would use rarely enough that a shortcut isn't necessary, and if I'm not using it frequently I would probably forget the shortcut anyway.}''
\end{quote}

There was nothing specific about \sysname{v2} encryption that participants disliked. The main issue participants mentioned for \sysname{v1} is its unintuitive UX because participants type the plaintext in the target application and see encrypted text in the target app and plaintext in a separate UI. This unintuitive UX requires a leap of faith from first-time users to trust the system and type their sensitive text in an untrusted app, hoping the app does not get access to their sensitive text.

\begin{quote}
    ``\textit{I just felt that was a bit scary at first, so I tried typing in the InfoGuard window itself but then I found out that's not how it works. I prefer the way how it works now to how I thought it worked, but it just took a leap of faith for me to be comfortable with it.}''
\end{quote}

Participants liked the intuitive UI of \sysname{v2}. Seeing the recipient's name while entering plaintext ensured they had selected the correct recipient.

\begin{quote}
    ``\textit{I like that the window is the focus of the screen, making it very clear as to where I should be looking and typing.}''
\end{quote}

\begin{quote}
    ''\textit {This one was cool because you could see who you were sending a message to while typing out. This could eliminate some errors with choosing the wrong person}''
\end{quote}

For decryption, some participants mentioned that it is cumbersome to highlight the exact encrypted text using the mouse. They would prefer just to double-click to highlight the encrypted text, which could be achieved using base62 encoding instead of base64 encoding.

\section{Discussion}

\paragraph{Developer API}
In addition to user-initiated E2EE, application-initiated encryption is possible by the developer using the socket API to trigger \sysname{'s} encryption mode.
This feature provides developers with flexibility and simplicity in integrating encryption into their applications without worrying about the complexity of the encryption process.


In our prototype, the Interceptor runs as a separate process that listens on a local socket. Applications use the socket API to trigger encryption instead of requiring the user to enable encryption mode and select the recipient.
The application can send a message to the socket to initiate encryption when a user clicks on the textbox that developers want to secure. The socket API calls typically take less than ~10 lines of code (see Listing  ~\ref{lst:develop_API}).

Application-initiated encryption assumes a trusted client because a compromised client can fail to invoke \sysname and display dummy overlays to capture sensitive content. When \sysname is invoked, it protects against client-side passive attacks because the client never gains access to plaintext. 

\begin{lstlisting}[caption={Code example to support E2EE encryption on a textbox by a developer}, label={lst:develop_API}]
var socket = new WebSocket("ws://localhost:5000");
document.getElementById("inputBox").addEventListener("click", function (event) {
    socket.onopen = function () { 
        var recipient = getMessageRecipient();
        socket.send(recipient);
    };
});
\end{lstlisting}

\paragraph{\textbf{Censorship}}
\sysname raises the bar on countries and organizations that censor encrypted communications. There is no app to ban based on well-known ports or protocol signatures. Blocking \sysname requires scanning the application data for encrypted text and blocking individual messages, which could not be done for TLS transmissions but could work on unencrypted protocols. However, the censor has to scan the application data actively and cannot block based on metadata. 

\paragraph{\textbf{Law enforcement}}
Law enforcement concerns sometimes lead to discussions about banning E2EE. Instead of an all-or-nothing approach, \sysname is a middle-of-the-road solution that allows a user to send a message and encrypt only the most sensitive data (e.g.~account numbers, SSNs, currency totals). It allows auditors to view the purpose and non-sensitive data in the message while safeguarding sensitive data. 

\paragraph{\textbf{User-space vs. kernel-space}} We implemented \sysname as a user-space application rather than a kernel module. Although a kernel module provides greater isolation from user-space vulnerabilities, a user-space application has several advantages. It is easier to compile for any platform with the necessary libraries and dependencies, making it more accessible for users in the short term. User space development is faster and simpler, with fewer memory allocation constraints and more straightforward debugging. Additionally, \sysname's isolation from the kernel ensures that early-stage adoption does not compromise the entire system due to code crashes or vulnerabilities.

\paragraph{\textbf{\sysname for other OSes and phones}}
Our primary goal was to demonstrate E2EE for desktop applications that have received relatively less interest from the security community but are as important as personal mobile chat applications. We presented our design and implemented our prototype for Linux-based systems, which can be easily extended to Windows and MacOS using the same interception technique. InfoGuard's keyboard interception can also be implemented for mobile applications since Android phones are Linux-based. However, displaying plaintext and encrypted text on different GUIs in mobile apps requires further research.

\paragraph{\textbf{Message storage}} We implemented \sysname using the Signal protocol, which provides forward secrecy by deleting the encryption/decryption key of sent/received messages. For permanent access, messaging clients often retain plaintext copies of the messages. \sysname can store messages locally under its permissions so the client app does not have access, but users can continue to read old encrypted messages.

\paragraph{\textbf{Searching encrypted text}}
\sysname aims to improve the usability of encryption by including a search function. To search the encrypted data, users must have access to encryption keys. Since encryption keys are deleted for forward secrecy, users cannot search the old encrypted messages through the application, even using homomorphic encryption. In addition, we cannot trust applications with plaintext and, therefore, cannot allow users to enter sensitive text for searching in the application's GUI. Therefore, $\sysg$ has to be used to search the locally stored sensitive data of any app where \sysname was used. However, users can search the unencrypted text as usual on the application's search interface.

\paragraph{\textbf{Key distribution and multi-device support}}
In systems lacking forward secrecy, such as PGP, users can synchronize their private key across all devices to decrypt messages on any device. 
However, syncing the private key will not work in systems like Signal since the protocol ratchets keys and deletes previous ones. Future research can explore approaches to provide multi-device support, such as group key management and shared cloud storage.

\section{Conclusion}

\sysname has a novel design to support user-controlled, application-independent encryption. It enables E2EE on any application without requiring changes to the application. \sysname encrypts text before it reaches the application, eliminating the client app's access to plaintext. It leverages visible encryption to help users better understand how their data is secured. A user study demonstrates that it is usable. Users trust it and are willing to adopt it. 

\sysname offers increased privacy for privacy-conscious users or organizations wanting to share sensitive data internally over third-party communication platforms. It has potential benefits for individuals living under censorship and surveillance. 








\bibliographystyle{plain}
\bibliography{reference}



\appendix

\section{\sysname Messages Flow}
\label{app:implementation_mess_flow}


\paragraph{\sysname's initialization}
In Unix-like operating systems, the \textit{/dev/input} directory contains device files for input devices such as keyboards and mice. The \textit{/dev/input/eventX} files are event files that correspond to input devices, with X being a number that identifies the specific device. A keyboard input file allows applications to receive raw keyboard events, including key presses and releases, without any processing or interpretation by the system. Generally, windowing systems such as Window System (X11) API or the Wayland protocol in Linux read keyboard input files, and applications receive keyboard input events from the windowing system.

To intercept the keyboard input, our system opens the relevant keyboard event file (e.g., \textit{/dev/input/eventX}) in a read-only, nonblocking mode. Then, it creates a virtual keyboard using \textit{uinput}, which it later uses to pass encrypted messages to the target application. The \textit{uinput} interface in Linux allows programs to create virtual input devices and send events to them through the /dev/uinput device file using the uinput API. Once created, events are sent to the virtual keyboard using the write system call, and the kernel receives them as if they were coming from a physical keyboard. 


\begin{figure*}
\centering
\includegraphics[width=0.95\textwidth]{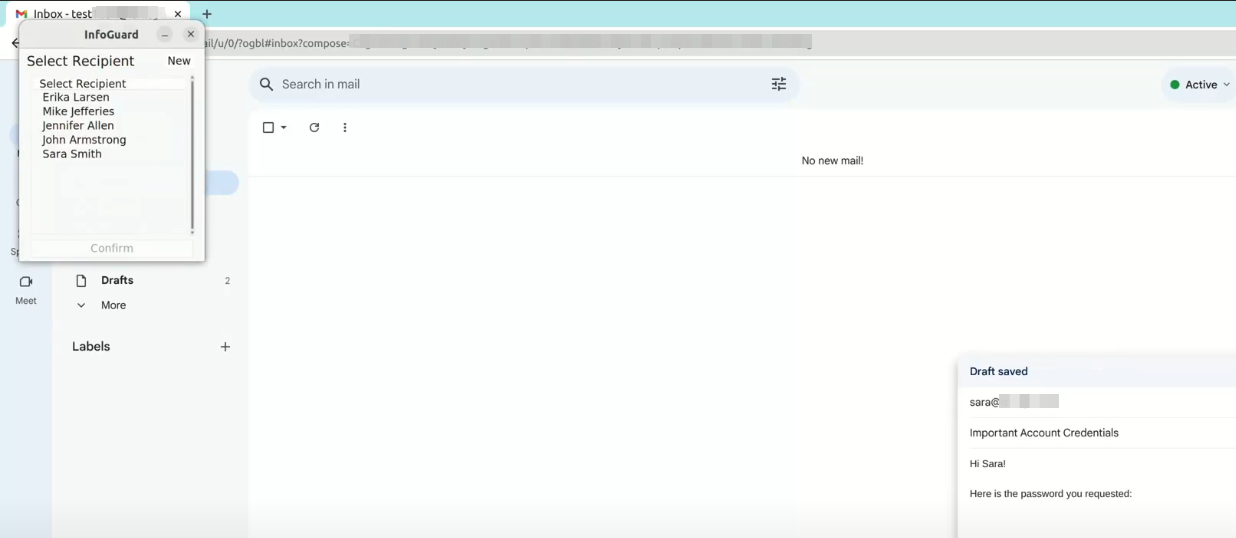}
\caption{\sysname{v1} encryption. The user first selects the recipient in $\sysg$ after they initiate encryption.}
\label{fig:infoGuard_v1_0}
\end{figure*}



\begin{figure*}
\centering
\includegraphics[width=0.95\textwidth]{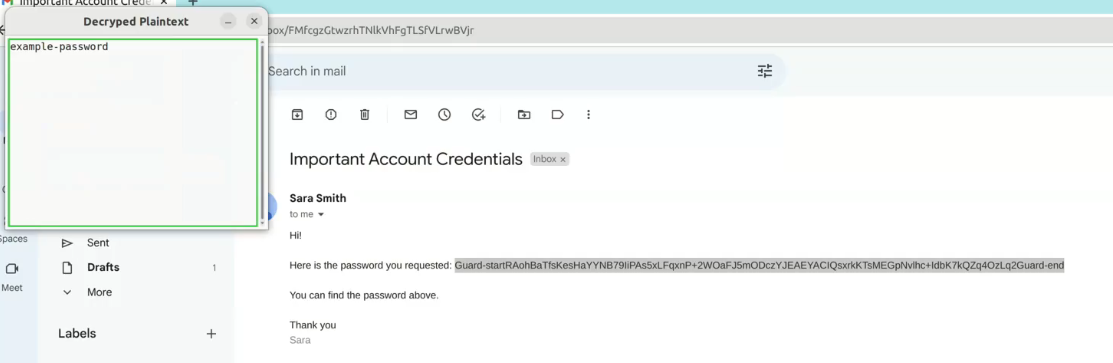}
\caption{\sysname decryption.}
\label{fig:infoGuard_decryption}
\end{figure*}

\paragraph{Waiting for user to enable encryption or decryption mode}
The interceptor daemon is in place and actively listening to the keyboard input file. To detect encryption and decryption shortcuts, Interceptor employs \textit{xkbcommon}, an open-source library designed for key event processing and keymap compilation in the X Window System protocol. Interceptor compiles a keymap using xkbcommon's keymap compiler and utilizes the resulting keymap to process key events. This entails setting up an event loop to obtain keyboard events and checking the keymap's state to identify whether the encryption or decryption shortcut has been pressed. When the daemon detects the correct sequence of keystrokes, the program switches to the appropriate mode, encryption or decryption.



\paragraph{Encryption mode initialization}
We use a state-of-the-art Signal E2EE protocol that provides forward secrecy and deniability. However, \sysname cannot use Signal protocol directly because Signal uses AES-OFB block cipher and expects the entire text to be typed before the encryption process starts. Interceptor expects to display the encrypted text in real-time as the user type, indicating to users that the input is being encrypted for the app they intend to use. The AES-OFB mode can be used as a synchronous stream cipher as it generates keystream blocks, which can be XORed character by character. However, OpenSSL does not provide an API to retrieve the keystream block instead it takes a 128-bit input and returns the encrypted text (plaintext\_block XORed keystream\_block). In our implementation to retrieve the keystream, we pass a block of 0s as input to OpenSSL for encryption and XOR the output with 0s again, which is equal to the keystream that OpenSSL used internally as XOR is a reversible function. Interceptor uses this 128-bit keystream to perform character-by-character encryption by XORing it with user input. When only 16 unused bits are left in the keystream, Interceptor encrypts an additional block of 0s (128 bits) and continues this till the user finish typing in the encryption mode. The C Signal library (\textit{libsignal-protocol-c}) is used for this implementation since the overall \sysname relies heavily on system library calls, particularly related to Libevdev.


 Interceptor uses the Libevdev library to grab exclusive access to the keyboard input file. It gains exclusive access to the keyboard, preventing other applications or processes from accessing it until the grab is released. It is achieved using the \textit{LIBEVDEV\_GRAB} flag in Linux.
Interceptor determines the appropriate location to place the final message \ie, the application's input textbox through which the user initiates encryption mode. It utilizes XGetInputFocus() from the Xlib library to keep track of the previous input application. 
Interceptor starts the Encryptor module used for encryption and decryption. Interceptor starts the $\sysg$. The $\sysg$ prompts the participants to either choose a recipient from available options or add a new recipient to encrypt their messages.


Once the recipient is selected, the user can begin typing in the application they wish to encrypt their message into, and the $\sysg$ displays the plaintext in real time. 


\paragraph{Encrypting the message}
As the user types the plaintext message, Interceptor forwards the ASCII characters to the $\sysg$, which dynamically displays the plain text characters that users type in the target application. Interceptor maintains a list of whitelist keys, such as backspace, shortcuts \textit{alt+F4}, \textit{Ctrl+A}, and \textit{Ctrl+C}, that it does not encrypt. These whitelisted keystrokes are passed to apps through the windowing system in plaintext.

Interceptor encrypts the user input character by character using the modified Signal protocol to allow for a stream cipher. 
To ensure all characters are printable in the target application's textbox, the Interceptor encodes the encrypted text with base64. Then, it wraps the base64 cipher text in a token format: "Guard-start{<cipher text>}Guard-end". This token allows the recipient an easy way to understand that the text is encrypted. The interceptor writes the base64 encoded encrypted text to the app using the virtual keyboard created during initialization. We determine that the minimum gap between two consecutive virtual keypresses is 0.00125 seconds before it corrupts the kernel's internal \textit{/dev/input} ring buffer, used to read virtual keyboard inputs by the application, leading to lost keystrokes. For efficiency, Interceptor waits for 0.3 seconds after the latest user's key presses before it starts encryption. It allows more efficiency and usability by waiting for the user to finish the block of words or sentences and encrypting them at once.

\begin{figure*}[hbtp!]
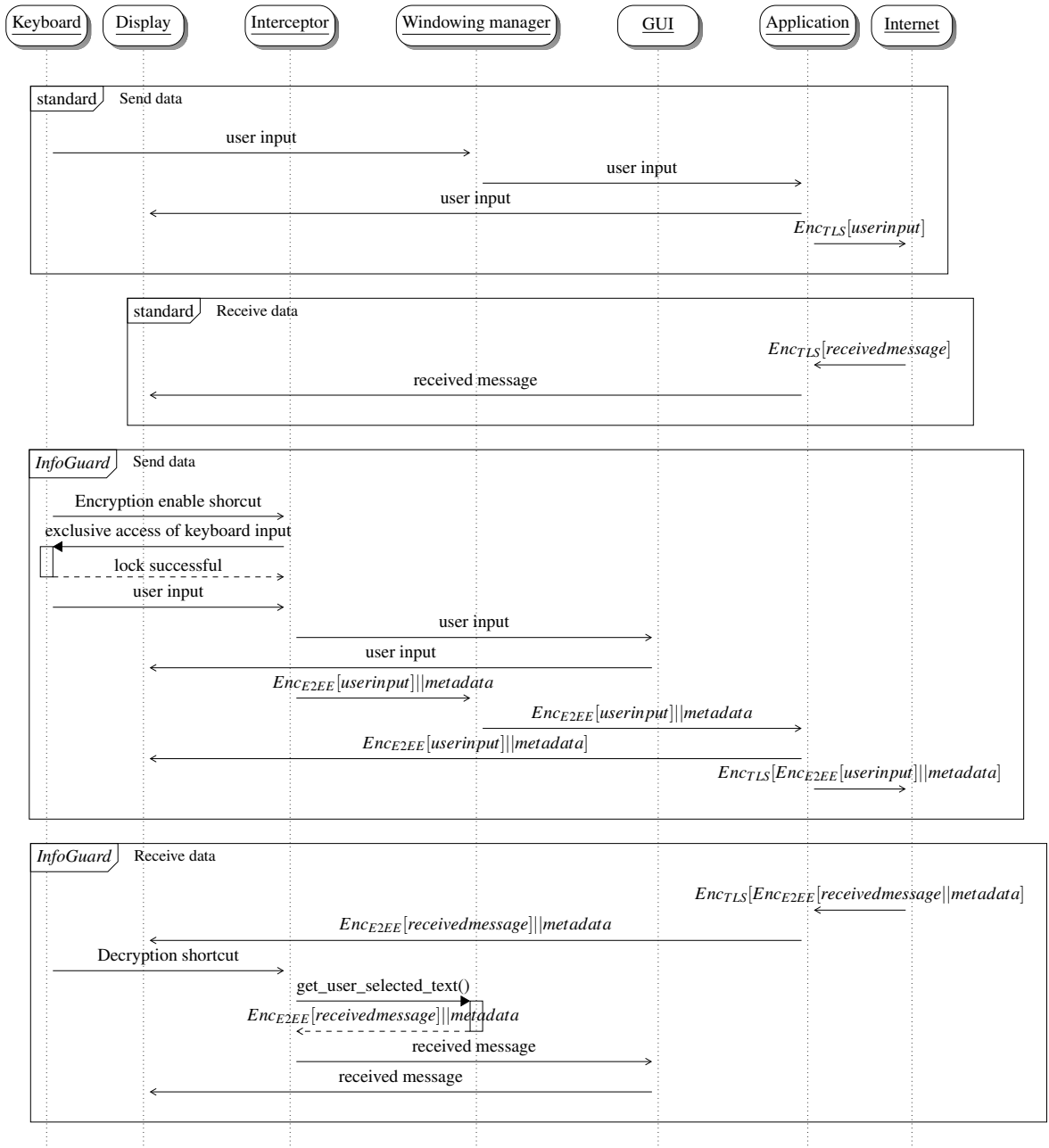

\begin{center}
\resizebox{0.9\textwidth}{!}{%
\begin{sequencediagram}
    \tikzset{inststyle/.append style={
        drop shadow={top color=gray, bottom color=white}, 
        rounded corners=2.0ex
    }
    }   
    \newinst{k}{Keyboard}
    \newinst[0.3]{d}{Display}
    
    \newinst[1.2]{i}{Interceptor}
    \newinst[1.2]{w}{Windowing manager}
    \newinst[1.2]{g}{GUI}
    \newinst[1.2]{a}{Application}
    \newinst[0.3]{in}{Internet}

    \begin {sdblock}{ standard }{ Send data }
        \mess{k}{user input}{w}
        \mess{w}{user input}{a}
        \mess{a}{user input}{d}
        \mess{a}{$Enc_{TLS}[user input]$}{in}
    \end{sdblock}

    \begin {sdblock}{ standard }{ Receive data }
    \mess{in}{$Enc_{TLS}[received message]$}{a}
    \mess{a}{received message}{d}
    \end{sdblock}
    

    \begin {sdblock}{ \sysname }{ Send data }
    \mess{k}{Encryption enable shorcut}{i}
    \begin{call}{i}{exclusive access of keyboard input}{k}{lock successful}
    \end{call}
    \mess{k}{user input}{i}
    \mess{i}{user input}{g}
    \mess{g}{user input}{d}
    
    \mess{i}{$Enc_{E2EE}[user input] || metadata$}{w}
    \mess{w}{$Enc_{E2EE}[user input] || metadata$}{a}
    \mess{a}{$Enc_{E2EE}[user input] || metadata]$}{d} 
    
    \mess{a}{$Enc_{TLS}[Enc_{E2EE}[user input] || metadata]$}{in}
    \end{sdblock}
    

\begin {sdblock}{ \sysname }{ Receive data }
      \mess{in}{$Enc_{TLS}[Enc_{E2EE}[received message || metadata]$}{a}
      \mess{a}{$Enc_{E2EE}[received message] || metadata$}{d}
      \mess{k}{Decryption shortcut}{i}
      \begin{call}{i}{get\_user\_selected\_text()}{w}{$Enc_{E2EE}[received message] || metadata$}
      \end{call}
      \mess{i}{received message}{g}
      \mess{g}{received message}{d}
\end{sdblock}

\end{sequencediagram}
}
\caption{\sysname message flow. Note, messages from the application/GUI to display are directly transmitted in this diagram for simplicity. In reality, messages to display are transmitted through the windowing manager.}
\label{fig:system}
\end{center}
\end{figure*}

\sysname also adds the metadata along with the cipher text on the target app's textbox, such as the message length, MAC, and Diffie-Hellman ratcheting keys, to the recipient along with every message. The total length of metadata is greater than 50 bytes \ie $50 + plaintextLength + floor(plaintextLength / 128) + 2*floor(messageNumber / 128)$.

\paragraph{Encryption termination}
When Interceptor detects the shortcut to disable encryption mode, it waits for the encryption function to finish and updates the latest encrypted text and metadata in the target application's textbox. Then, Interceptor adds the ciphertext hash and plaintext mapping to the local cache. This local storage allows the user to again see the plaintext messages through the ciphertext message later. Without this, the sender will not be able to see a message after sending it, and the recipient will not be able to see it after the first decryption. To achieve forward secrecy, Signal deletes the previous keys. Interceptor cleans up the Signal encryption context. In particular, the Signal keys are appropriately ratcheted, the one-time pad is zeroed out and freed, and the underlying cryptographic structures are freed. Interceptor releases its exclusive grab of the keyboard input file, thus allowing other applications to resume reading from it. Interceptor kills the $\sysg$.



\paragraph{Decryption on the receiver end} 
In \sysname, we incorporate user involvement in the decryption process to achieve translucency, which helps to increase trust in our system. Interceptor actively listens for the encryption/decryption shortcut, which the user can trigger to view the decrypted message. When users receive an encrypted text, they only see the ciphertext in the application. To decrypt the message, the user selects the encrypted text and presses the decryption shortcut, which is \textit{Ctrl+Alt+U}.

Interceptor retrieves the selected text from the application using Xlib, which enables any application to retrieve selected text using the PRIMARY selection. Once the selected ciphertext is obtained, Interceptor pops up a graphical user interface (GUI) window for the user to choose the message's sender. It then uses the key session with the selected sender to decrypt the received encrypted message. Similar to the encryption mode, Interceptor pops up another $\sysg$ window to show the plain text to the users.


\paragraph{Developer's API}
Interceptor exposes a socket to all applications that allow them to trigger encryption. Application developers can notify the Secure Encryption ID (\sysname) through a socket to initiate encryption. To detect when a user starts typing in a textbox that needs to be secured, developers can use listener functions such as "onfocus" in JavaScript and "textbox.bind" in Tkinter (for Python).

It should be noted that the scope of this design does not include protection against attacks on the client app itself that may disable the \sysname call. However, it is important to highlight that such attacks are possible due to vulnerabilities in the application, not the \sysname system. \sysname ensures that if a user wants to start encryption, the application cannot access plaintext data even if it is actively trying. However, if the user explicitly chooses not to use \sysname, the responsibility for encryption falls on the application developer. In such cases, the level of security provided by the application's encryption mechanism is as secure as the client end of the application. \sysname runs with sudo privilege and therefore other non-sudo apps cannot delete or impersonate the socket API to perform denial of service attacks on other apps.

\begin{table}[tbp]

    \caption{E2EE coverage}
    \label{table:app_analysis}
    \begin{center}
    \begin{tabular}{|p{7cm}| p{0.7cm} |}
        \hline
         \textbf{Apps} & \textbf{E2EE}\\
        \hline
        \textbf{Personal} & \\
        \hline
        WhatsApp, Viber, Signal, Snapchat & \yes\\
        \hline
        Instagram, Facebook Messenger, Imo, Line, Telegram 
        & \sometimes \\
       \hline
        WeChat, Pinterest, TikTok, Weibo, QQ, Douyin, Reddit, Quora, Kuaishou, Qzone, Picsart, Likee, Twitch, Stack Exchange, Tieba & \no\\
        \hline
        
        \textbf{Business} & \\
        \hline
        Troop Messenger, Brosix & \yes \\
        \hline
         Microsoft Teams & \sometimes\\
       \hline
        LinkedIn, Clariti, Zulip, Mattermost, Asana, Trello, Flock, Chanty, Pumble, Jostle, Workvivo, Bitrix24, Workplace, Pronto, Twist, Basecamp, Zoho Cliq,  Crew, Fleep & \no\\
        \hline
        \textbf{Both} & \\
        \hline
        Slack, Google Chat, Discord, WeChat, Twitter, Youtube & \no  \\


        \hline
    \end{tabular}
    \end{center}
    \begin{tabular}{l}
    \end{tabular}
    \yes = provide E2EE, \sometimes = has an optional E2EE, \no = no E2EE 
    
\end{table}

\section{\sysname Implementation}
\label{app:implementation_details}
We built proof-of-concept prototypes of both versions of \sysname to demonstrate their feasibility, evaluate their performance, and conduct user studies. We first describe \sysname{v1} in detail and then describe additional components of \sysname{v2}.

\subsection{\sysname{v1}}


Our implementation of the Interceptor is a native C/C++ application that runs the $\sysg$. We employed the Tkinter Python library for Tcl/Tk to develop a user-friendly and adaptable design with cross-platform compatibility for $\sysg$. We utilized the Signal C library (libsignal-protocol-c)~\cite{signalclibrary} to implement forward secrecy encryption and decryption, which we modified to support a stream cipher. 

\paragraph{Interceptor}

\begin{enumerate}
    \item \textit{Listens for encryption/decryption shortcuts}: In Unix-like operating systems, the \textit{/dev/input} directory contains device files for input devices like keyboards, with \textit{/dev/input/eventX} files corresponding to specific devices. \sysname intercepts keyboard input by opening the relevant event file in read-only mode. It detects encryption and decryption shortcuts using \textit{xkbcommon}, a library that compiles a keymap and processes keyboard events by checking the keymap's state. The program switches to the appropriate mode when detecting the correct keystroke sequence.
    
    \item \textit{Input interception}: Interceptor uses the Libevdev library to gain exclusive access to the keyboard input file by setting the "LIBEVDEV\_GRAB" flag while accessing it, which prevents other applications or processes from accessing it until the grab is released.

    \item \textit{Plaintext transmission to $\sysg$}: We use sockets to transmit plaintext to the $\sysg$ process from the Interceptor. In \sysname, Interceptor runs as a separate sudo user with a unique UID and GID, allowing for separate permissions and access controls. This approach protects against accessing inter-process communication. Furthermore, it prevents potential attacks that exploit vulnerabilities in the user's system, including the possibility of an attacker using \textit{ptrace} to attach to the process and dump memory. By default, some distributions have ptrace\_scope set to 0, which allows any process under the same user to attach to any other process under the same user. Running the Interceptor as a separate user mitigates this risk.

    \item \textit{Real-time encryption and transmission}: 
    Interceptor uses the Signal protocol for end-to-end encryption, forward secrecy, and deniability. The Signal encryption (AES-OFB) was modified from a block cipher to a stream cipher to enable real-time encryption. To retrieve the keystream block, Interceptor passes a block of 0s as input to OpenSSL for encryption and XORs the output with 0s again. The resulting keystream is used to perform character-by-character encryption by XORing it with user input. The interceptor encrypts an additional block of 0s (128 bits) when only 16 unused bits are left in the keystream and continues until the user finishes typing in the encryption mode. The C Signal library is used for the implementation.
    To transmit data to the app, Interceptor creates a virtual keyboard using the \textit{uinput} interface in Linux. To ensure that all characters are printable, the Interceptor encodes the encrypted text with base64 and wraps it in a \textit{Guard-start<encrypted text>Guard-end} format. 
    It writes the base64-encoded encrypted text with metadata to the target app using the virtual keyboard.

    \item \textit{Decryption}: When a user wants to decrypt an encrypted message in \sysname, they select the ciphertext in the application and press the decryption shortcut. Interceptor uses Xlib's PRIMARY selection to retrieve the selected text from the application and then performs decryption using the modified Signal protocol. The decrypted message is then transmitted to the $\sysg$, following the same process described in the previous item.

    \item \textit{Developer API}: Interceptor provides a socket that allows applications to trigger encryption. Application developers can use listener functions to detect when a user starts typing in a textbox that requires secure encryption. However, it is important to note that \sysname does not protect against attacks on the client app itself, which may disable the \sysname call. In such cases, the responsibility for encryption falls on the application developer. \sysname runs with sudo privilege, making it impossible for non-sudo apps to delete or impersonate the socket API for denial of service attacks on other apps.
\end{enumerate}

\paragraph{$\sysg$}
The Interceptor starts the $\sysg$ as a child process. As shown in Fig~\ref{fig:infoGuard_v1}, $\sysg$ displays the plaintext dynamically as it is received from the Interceptor. Furthermore, $\sysg$ has a setting option for the user to configure properties such as encryption algorithms.


\paragraph{Secure implementation ideas}
When implementing \sysname, it is crucial to consider the security risks that may be introduced during implementation to ensure its security. The attacks based on implementation are out of the scope of this paper; however, we discuss the two most common implementation attacks that pose a high risk: shared object injection and Python library substitution. Shared object injection involves an attacker setting the \textit{LD\_PRELOAD} environment variable to replace essential libraries with malicious ones. Python library substitution consists of an attacker overwriting userspace installed dependencies with malicious modules. This could result in the interception of sensitive information without any special permissions. One possible mitigation is bundling all dependencies together, ensuring all required libraries are included in a self-contained environment.
 
Also, the implementation of \sysname needs to ensure that private keys and decrypted messages are stored in a directory owned by the user who runs the Interceptor and has restricted permissions.

\subsection{\sysname{v2}}
Our implementation framework remains the same as \sysname{v1} with some alterations to the user input process. The interceptor component actively monitors encryption and decryption shortcuts and triggers the appearance of $\sysg$ when encryption is activated. In \sysname{v2}, $\sysg$functions is an overlay, allowing users to input text directly into a designated secure textbox. Once the user has input their sensitive message, the plaintext is transmitted to the Interceptor component. The Interceptor then proceeds to perform block cipher encryption using AES-OFB mode. The virtual keyboard feature copies the encrypted text to the clipboard and the content is automatically pasted into the target application, ensuring a seamless and secure user experience.

\section{Potential usage of \sysname}
We analyzed the top 100 Alexa websites to find which of these platforms \sysname can be used to secure user-to-user communication. Here are the following applications with descriptions of usage:

\begin{enumerate}
    \item Google.com: email 
\item youtube.com: 
\item Amazon.com: ‘Amazon prime video watch party’ chat 
\item yahoo.com: email
\item facebook.com: messages/ posts shared to specific friends/groups
\item zoom.us: chat 
\item force.com: The Text Messaging Service to Stay Connected with the customers
\item reddit.com: Reddit chat
\item Office.com: Any sensitive content in the document that is supposed to be shared only with a person/group securely
\item Bing.com: 
\item shopify.com: Website owners can enable encryption such that whatever chat messages their customer sends to them are E2EE and cannot be seen in between by the server they are running the service on. 
\item eBay.com: chat with seller
\item instagram.com: Chat/ post share for specific group
\item live.com: Emails/ sensitive documents. 
\item Chase.com: communication with customer service.
\item microsoft.com: Microsoft Teams chat
\item Netflix.com: Netflix watch party chat
\item Microsoftonline.com:  
\item zilloq.com: customer service chat
\item Intuit.com: Turbotax’s customer care chat
\item Instructure.com: communication between student and instructors/teachers.
\item Twitch.tv: 
\item Linkedin.com: chat 
\item Twitch.tv: chat
\end{enumerate}


\end{document}